# Screening Highly Effective Electrocatalyst for $N_2$ Reduction on Anatase $TiO_2$(101) Surface: By Tuning Electronic and Geometry Features

Tongwei Wu,[†] Yanning Zhang[*,†]

[†]Institute of Fundamental and Frontier Sciences, University of Electronic Science and Technology of China, Chengdu 610054, Sichuan, China

**ABSTRACT:** Industrial-scale $NH_3$ production is mainly produced by the Haber−Bosch process, but it suffers from intensive energy consumption with serious $CO_2$ emission. Electrochemical $N_2$ reduction reaction (NRR) is now used for energy-saving $NH_3$ synthesis, which needs highly-efficient electrocatalysts for $N_2$ activation. In this work, we have computationally studied what are main factors for affecting $N_2$ molecule activation and NRR process on $V_{O2c}$-$TiO_2$ (101) surface by changing its structures or electronic features. Furthermore, a promising NRR electrocatalyst is successfully predicted.

## INTRODUCTION

$NH_3$ is vital to life on earth. In the early 1900s the invention of Haber−Bosch process ($N_2$ + $3H_2$ → $2NH_3$) has been used to industrially reduce $N_2$ to $NH_3$ applied in agriculture, fertilizer and pharmaceutical productions.[1-4] However, it is high energy demand arisen from the high reaction temperatures and pressures, consuming 1 to 2% of the world's annual energy output.[5-7] The process uses the $H_2$ as feeding gas obtained by the reformation of fossil fuels, which greatly increases $CO_2$ emissions. Electrochemical $N_2$ reduction reaction (NRR) using proton from water as the $H_2$ source can be powered by renewable electricity from solar or wind sources at mild conditions, emerges as an attractive alternative in an environment-friendly and sustainable manner.[8-9] Nevertheless, the strong and non-polarizable N≡N bond seriously restricts its activation, and thus a highly active catalyst for NRR is desired.

Experimentally, various electrocatalysts, such as noble metals (Au, Pd etc.), non-noble metal ($BiVO_4$, $MoS_2$ and $Cr_2O_3$ etc.), and metal-free ($BC_3$, $B_4C$ etc.), are recently studied for NRR.[10-19] It suggests that catalytic activity and selectivity for NRR come from the introduction of active species/sites. For example, Yao et al reported that $BiVO_4$ with oxygen vacancy is capable to promote NRR; Sun et al found that $MoS_2$ exhibits good NRR activity while the Mo-edge plays the key role to polarize the $N_2$ molecules. Note that the adsorption and activation of $N_2$ is the sufficient and necessary conditions for NRR. Therefore, what are main factors for affecting $N_2$ molecule activation is urgently expected for the accurate design of $N_2$ reduction electrocatalyst.

Due to abundance, nontoxicity, and high stability, $TiO_2$ have been widely studied in many applications, including photo-catalysis water splitting, $CO_2$ reduction to chemical fuel, and electron conductors in solar cell etc.[20] Recently, experiments reported that modified-$TiO_2$ materials through various strategies such as oxygen-defect, heteroatom-doping and rGO-hybridization, show good NRR performance.[21-25] With collaborations with experiments, our and other calculation results clearly demonstrate that the oxygen vacancy on $TiO_2$ (101) surface is responsible for catalyzing NRR. Based on these results, we wonder if the systematical investigation on this surface by changing its structures or electronic features could open up an universal guideline to the rational design of transition metal oxide and other compound electrocatalysts for artificial $N_2$ fixation.

## RESULTS AND DISCUSSION

We select $V_{O2c}$-$TiO_2$ (101) surface decorated with various number of H or Cl atoms to obtain different $Ti^{3+}$ concentrations (θ), as shown in **Figure 1 and Figure S2**. Therefore, we define the θ as the ratio of Ti atoms with excessive electrons over the total relaxed Ti atoms. Here we modify the number of $Ti^{3+}$ by introducing Cl atoms. Firstly, $N_2$ prefers to adsorb on $Ti_{4c}$ site by end-on coordination on $V_{O2c}$-$TiO_2$ (101) surface in **Figure 1a**. With the increasing of θ, $N_2$ adsorption energy decreases from -0.628 to -0.467 eV, the N≡N bond length increases from 1.112 to 1.118 Å and magnetic moment ($\mu_B$) increases from 0 to 2 $\mu_B$. Importantly, the bader charge analysis shows that $N_2$ molecule is activated by the electron injection from 0 to 0.08 $e^-$ as increasing θ. The comparison of PDOS of $V_{O2c}$-$TiO_2$ (101) surface before and after $N_2$ adsorption shows an obvious hybridization between Ti-3d and *$N_2$-2p orbitals below the Fermi energy, termed π backdonation, in **Figure S3**. When one $Ti^{3+}$ is induced by adsorbing one H atom on clean TiO$_2$ (101) surface (top panel in **Figure S4**), we can see that excessive electronic is localized on outermost surface Ti$_{5c}$ atom and then N$_2$ prefers to adsorb on it by end-on coordination. We compare V$_{O2c}$+Cl- and H-decorated TiO$_2$ (101) surfaces which have the same number and distribution of Ti$^{3+}$, and the result shows that PDOS of Ti$^{3+}$ has no difference in **Figure S4**. However, the bond length and charge of N$_2$ are unchange on H-decorated TiO$_2$ (101). It demonstrates that the O-vacancy not only gives a good space for N$_2$ adsorption, but induces charge accumulations to its Ti neighbors. The latter changes the electronic and structure features of N$_2$, making it more active for the following NRR process.

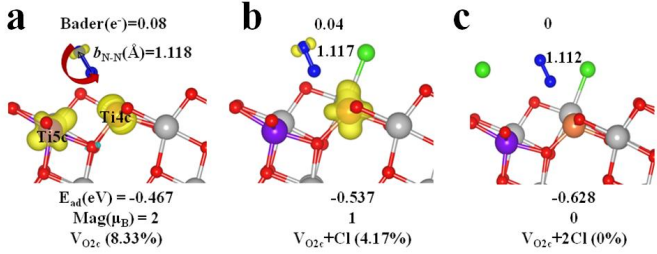

**Figure 1.** N$_2$ adsorption and activation on V$_{O2c}$-TiO$_2$ (101) surface with various θ (0, 4.17 and 8.33%). Yellow isosurface represents topological structure of Ti$^{3+}$. Ti, gray; Ti$^{3+}$ (Ti$_{4c}$), orange; Ti$^{3+}$ (Ti$_{5c}$), purple; Cl, green; O, red; N, blue.

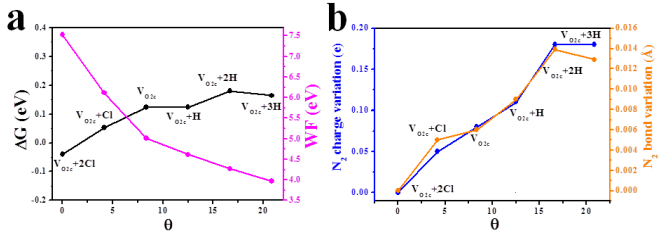

**Figure 2.** (a) Free energy of N$_2$ adsorption and work function on V$_{O2c}$-TiO$_2$ (101) surface and (b) the corresponding of bond and charge variations of N$_2$ as θ.

Then we calculate the free energy of N$_2$ adsorption on V$_{O2c}$-TiO$_2$(101) with various θ, and see the value on the surfaces linearly changes from negative to positive with increasing θ (**Figure 2a**), implying that N$_2$ adsorption seems to decrease. Moreover, work function (WF) linearly decreases as increasing θ on V$_{O2c}$-TiO$_2$(101) surface in **Figure 2a**. In general, N$_2$ is activated by forming π back-bonding and its electronic variation is a strong evidence of π back-bonding. Therefore, the bond elongation and electronic transition of *N$_2$ intermediate could judge whether N$_2$ is activated. Subsequently, we use the bond and charge variations of N$_2$ as parameters, and plot their changes as the θ value, and results show that it increases with increasing θ, as shown in **Figure 2b**. We also plot their changes as WF and results show the free energy of N$_2$ adsorption linearly changes from positive to negative as increasing WF in **Figure 3a**. Both the bond and charge variations of N$_2$ present reduction as increasing WF in **Figure 3b**. In addition, we compare V$_{O2c}$-TiO$_2$(101) surface mixed with other decorated species including Cl, H, Ti$_{int}$ and V$_{O3c}$ to V$_{O2c}$-TiO$_2$(101) surface decorated by the number of various H or Cl atoms (**Figure 3c and d**), and a similar rule is obtained, demonstrating that the aforementioned results are reliable. Overall, these data clearly indicate that N$_2$ adsorption and activation can be tuned by changing the WF or θ on V$_{O2c}$-TiO$_2$ (101) surface. In experiment, heteroatom-doping or other approaches can be used to tune them.

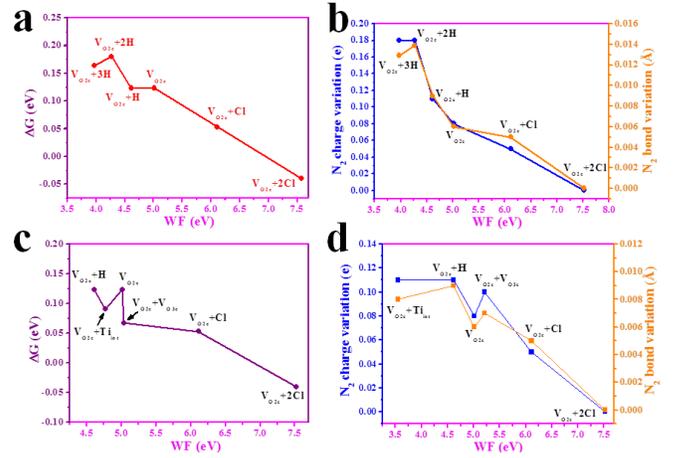

**Figure 3.** (a) Free energy of N$_2$ adsorption on V$_{O2c}$-TiO$_2$ (101) surface and (b) corresponding of bond and charge variations of N$_2$ as WF. (c) Free energy of N$_2$ adsorption on TiO$_2$ (101) surface decorated with either V$_{O2c}$ or V$_{O3c}$, as well as the Cl, H, Ti$_{int}$ or the mixing of them and (d) corresponding of bond and charge variation of N$_2$ as WF.

Subsequently, the whole NRR process on V$_{O2c}$+Cl and V$_{O2c}$-TiO$_2$ (101) surface are considered. We firstly calculate V$_{O2c}$-decorated TiO$_2$ (101) surface for NRR in **Figure 4a**, and then the various possible sites of first hydrogenration process for NRR are further considered, as shown in **Figure S5**. Results show that first hydrogenation process prefers to adsorb on surface oxygen (O$_{2c}$) to form OH bond rather than *NNH intermediate, except for V$_{O2c}$+3H-TiO$_2$ (101) surface. Hence, we suppose that V$_{O2c}$-decorated TiO$_2$ (101) surface partially reduced by hydrogen atom is meaningful for NRR process. On the other hand, no regardless of V$_{O2c}$ or V$_{O2c}$+3H-TiO$_2$ (101) surface, the NRR process is the alternating and mixed pathways, and the distal pathway is neglected because it is terminated at *NNH$_3$, as shown in **Figure 4, S6-7**. An obvious difference between them through the alternating and mixed pathways are *NNH to *NNH$_2$ process, namely, the V$_{O2c}$-TiO$_2$ (101) surface presents downhill pathway while V$_{O2c}$+3H-TiO$_2$ (101) surface is opposite. For V$_{O2c}$+Cl-TiO$_2$ (101) surface that bis-Ti$^{3+}$ neighbor to V$_{O2c}$ are decreased to single Ti$^{3+}$ in **Figure 1b**, N$_2$ to NH$_3$ conversion is terminated at *NNH$_3$ and *NH$_2$NH$_2$ intermediates through distal and alternating pathways. It is importantly noted that *NH$_2$NH$_2$ is adsorbed on V$_{O2c}$+Cl-TiO$_2$ (101) surface by end-on coordination with Ti$^{3+}$ (Ti$_{4c}$), but it is adsorbed on V$_{O2c}$-TiO$_2$ (101) surface by side-on coordination with bis-Ti$^{3+}$ (Ti$_{4c}$ and Ti$_{5c}$), in **Figure S8**. Above date suggest that the bis-Ti$^{3+}$ neighbor to V$_{O2c}$ is responsible for NRR performance and the breaking of N-N bond in *NH$_2$NH$_2$ is a vital reaction step.

Based on aforementioned rules, we predict NRR process on V$_{O2c}$+Ti$_{int}$-docorated TiO$_2$ (101) surface (WF = 4.77 eV)

which is believed to be the active species in various surface catalysis, as shown in **Figure 4c and S9**. The calculation results show that NRR process experience three pathways including of the alternating, distal and mixed, and the rate-determining step appears at the desorption of second $NH_3$ with a free change of 0.77 eV. Surprisingly, the $*NNH_3$ intermediate is also broken as $*N$ and $*NH_3$. It suggests that $V_{O2c}+Ti_{int}$-docorated $TiO_2$ (101) surface may possess more active for NRR.

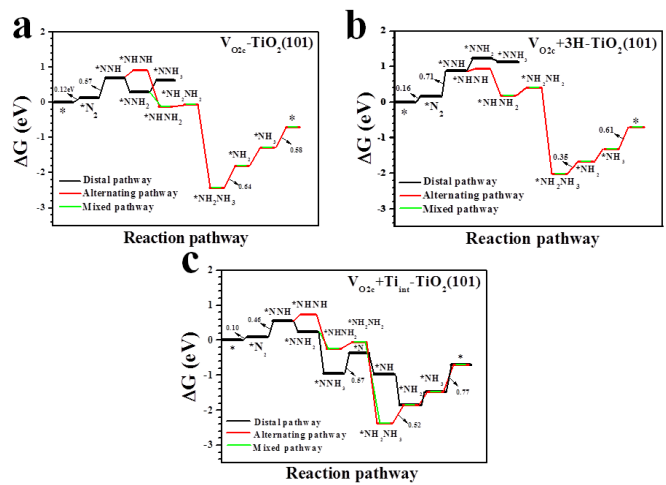

**Figure 4.** (a-c) Free energy diagram of NRR on $V_{O2c}$, $V_{O2c}+3H$ and $V_{O2c}+Ti_{int}$ - decorated $TiO_2$ (101) surface.

We also consider the performance of HER (a side reaction of NRR) for all surfaces from θ =4.17% to 20.8%, and results show that $V_{O2c}+2H$ (16.66%) and $V_{O2c}+3H$ (20.8%) -decorated $TiO_2$ (101) surface have an relatively worse HER performance, particularly $V_{O2c}+3H$-$TiO_2$ (101) surface with ΔG = 2.15 eV, as shown in **Figure 5a**. In addition, the performance of HER on $TiO_2$ (101) surface decorated with either $V_{O2c}$ or $V_{O3c}$, as well as the Cl, H, $Ti_{int}$ or the mixing of them are calculated in **Figure 5b**. Among these surface, $V_{O2c}+Ti_{int}$-docorated $TiO_2$ (101) surface exhibit worse HER performance with ΔG = -0.277 eV. These results demonstrate that NRR on $V_{O2c}+3H$ and $V_{O2c}+Ti_{int}$-docorated $TiO_2$ (101) surface may exhibit better NRR activity and selectivity.

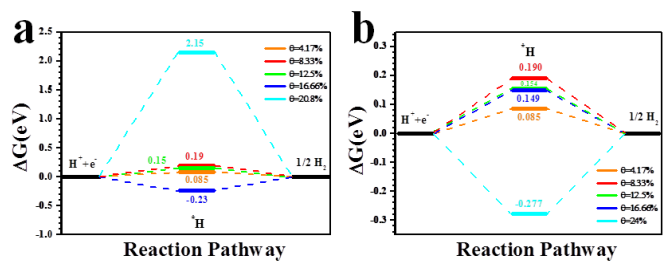

**Figure 5.** (a) Free energy diagram of HER on $V_{O2c}$-$TiO_2$ (101) surface decorated the number of difference H or Cl atoms. (b) Free energy diagram of HER on $TiO_2$ (101) surface decorated with either $V_{O2c}$ or $V_{O3c}$, as well as the Cl, H, $Ti_{int}$ or the mixing of them.

## CONCLUSION

In summary, using first-principles calculations, we have computationally studied NRR activity on $TiO_2$ (101) surface by changing electronic and structure features. Our calculation results show that the existing of $V_{O2c}$ on $TiO_2$(101) surface activates $N_2$ molecule, and its adsorption and activation can be tuned by tuning the WF or θ value. Furthermore, the synergistic role of bis-$Ti^{3+}$ induced by $V_{O2c}$ is responsible for NRR process on $TiO_2$(101) surface, in which the breaking of N-N bond in $*NH_2NH_2$ is a vital reaction step. Based on above direction, we have successlly predicted $V_{O2c}$-$TiO_2$ (101) surfaces partially reduced by three H atoms and decorated by $Ti_{int}$ atom for NRR. These results may open up an universal guideline to the rational design of transition metal oxide or other compound electrocatalysts for artificial $N_2$ fixation.

## ASSOCIATED CONTENT

### Supporting Information

Calculation details; Structure diagram; Topological structure; DOS; Atom configurations. This material is available free of charge via the Internet at http://pubs.acs.org.

## AUTHOR INFORMATION

### Corresponding Author

*E-mail: yanningz@uestc.edu.cn (Y.Z.);

### Notes

The authors declare no competing financial interest.

## ACKNOWLEDGMENT

This work was supported by the National Natural Science Foundation of China (No. 11874005).

# Table of Contents Graphic

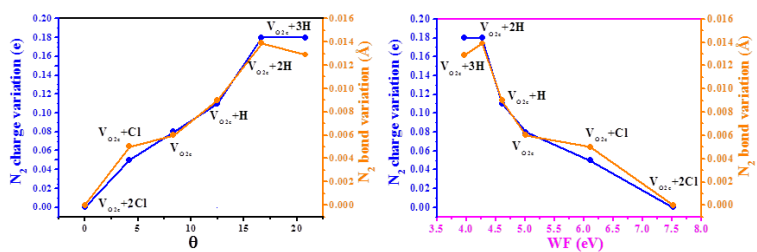

We have computationally demonstrated that the $N_2$ adsorption and activation can be tuned by changing the WF and $\theta$ on $V_{O2c}$-$TiO_2$ (101) surface. Moreover, the bis-$Ti^{3+}$ neighbor to $V_{O2c}$ is responsible for NRR performance and the breaking of N-N bond in $*NH_2NH_2$ is a vital reaction step.

# Supporting Information

**Computation and model details**

Spin-polarized density functional theory (DFT) calculations were performed by using the plane wave-based Vienna ab initio simulation package (VASP).[1,2] The generalized gradient approximation method with Perdew-Burke-Ernzerhof (PBE) functional was used to describe the exchange-correlation interaction among electrons.[3,4] The van de Walls (vdW) correction with the Grimme approach (DFT-D3) was included in the interaction between single molecule/atoms and substrates.[5] The energy cutoff for the plane wave-basis expansion was set to 500 eV and the atomic relaxation was continued until the forces acting on atoms were smaller than 0.01 eV $\text{Å}^{-1}$. The Brillouin zone was sampled with $3 \times 3 \times 1$ Gamma-center k-point mesh, and the electronic states were smeared using the Fermi scheme with a broadening width of 0.1 eV. The $TiO_2$(101) surface was modeled with a $2 \times 1$ slab in the lateral plane separated by 15 Å of vacuum in between to avoid the interaction between the slab and its period images. The free energies of the reaction intermediates were obtained by $\Delta G = \Delta E_{ads} + \Delta ZPE - T\Delta S$, where $\Delta E_{ads}$ is the adsorption energy, ZPE is the zero point energy and S is the entropy at 298 K. In this study, the entropies of molecules in the gas phase are obtained from the literature.[6]

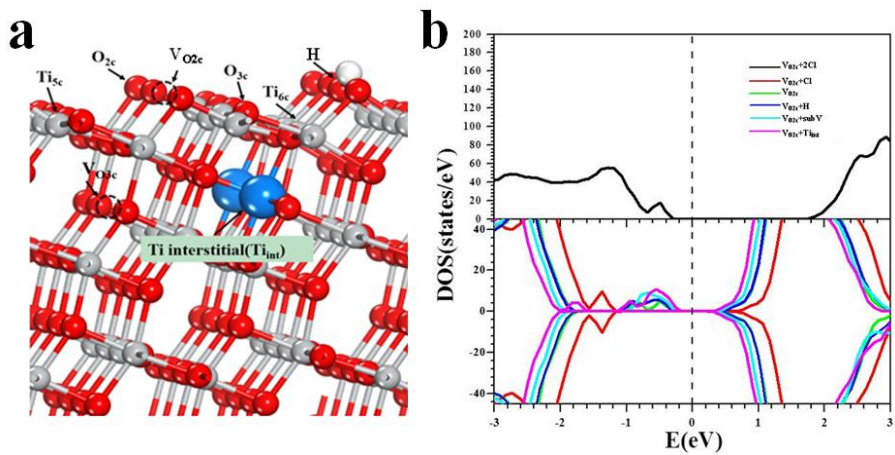

**Figure S1**. (a) The structure diagram of anatase $TiO_2$ (101) surface. H, $V_{O2c}$, $V_{O3c}$ and $Ti_{int}$ represent surface hydrogen atom, surface oxygen vacancy, subsurface oxygen vacancy and interstitial Ti atom. The red, gray and blue represent O, Ti and interstitial Ti atoms respectively. (b) Density of states of $TiO_2$ (101) surface decorated with either $V_{O2c}$ or $V_{O3c}$, as well as the Cl, H, $Ti_{int}$ or the mixing of them.

$TiO_2$ (101) surface has five-coordinated $Ti_{5c}$ and 2-coordinated $O_{2c}$ on the outermost surface, and $Ti_{6c}$ and $O_{3c}$ are on the deeper layers, as shown in **FigureS1**a. We then decorate the $TiO_2$ (101) surface with either surface or sub-surface oxygen vacancies ($V_{O2c}$ and $V_{O3c}$), as well as the existing of Cl atom, hydrogen atom (H), Ti interstitial atom ($Ti_{int}$) or the mixing of them. Thus, we manually introduce or eliminate some electrons in the systems. Then, some fully occupied impurity states appear just below the Fermi level in the band gap in **Figure S1b**. The integration on the spin density in this area fully come from the introduced excess electrons. The partial charge further confirms these excess electrons are localized on Ti atom.[7]

The $Ti^{3+}$ concentration is calculated using the following equation:

$$\theta = n/N \qquad (1)$$

Where $n$ is the number of $Ti^{3+}$ and N is the total number of Ti atoms over the total relaxed Ti atoms.

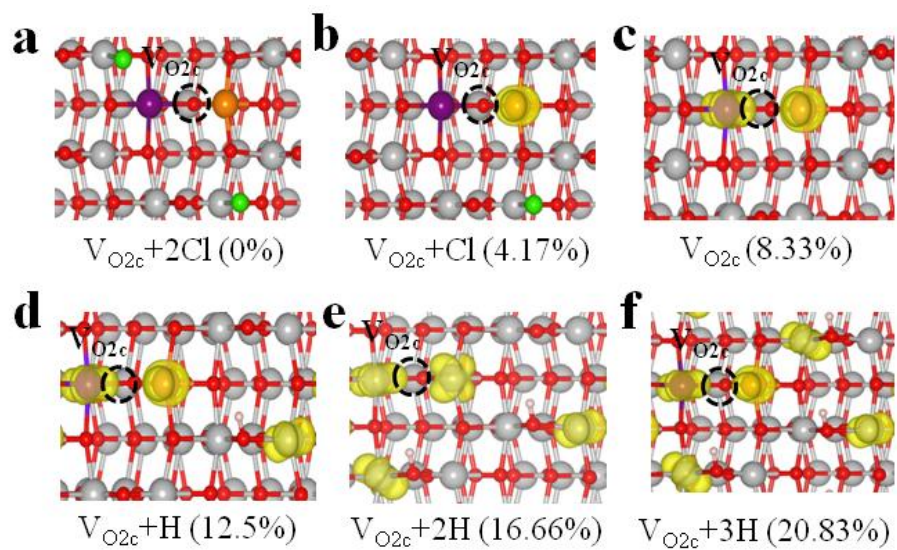

**Figure S2.** $V_{O2c}$-$TiO_2$ (101) surface with various θ. Yellow isosurface represents topological structure of $Ti^{3+}$.

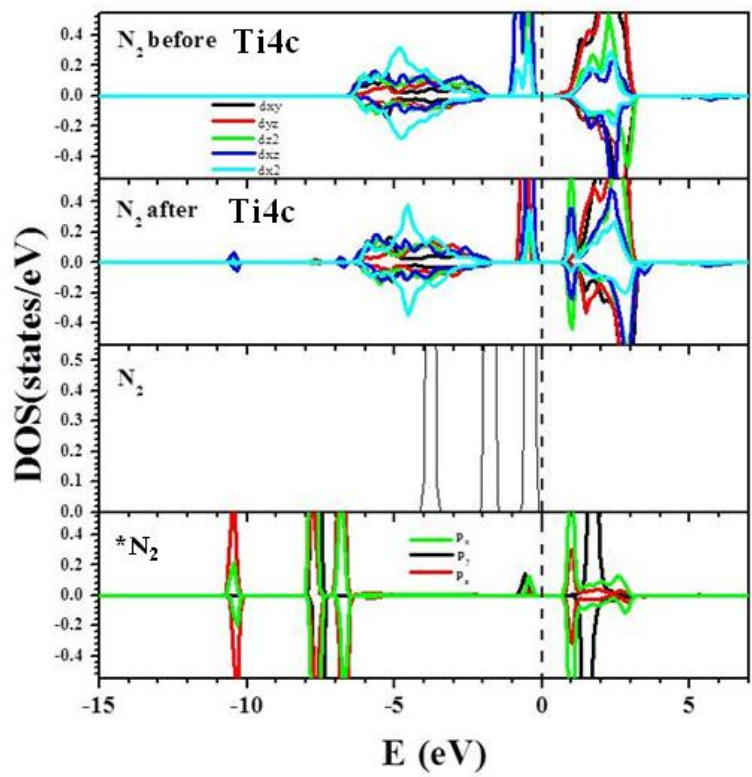

**Figure S3.** DOS of before and after $N_2$ adsorption on $V_{O2c}$-$TiO_2$ (101) surface.

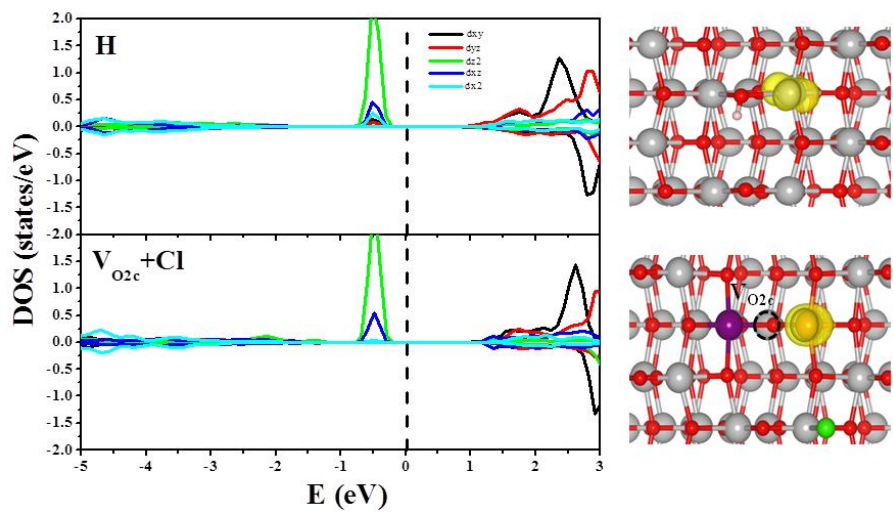

**Figure S4.** DOS of $V_{O2c}$ and $V_{O2c}$+Cl-decorated $TiO_2$ (101) surface and corresponding of topological structure of $Ti^{3+}$.

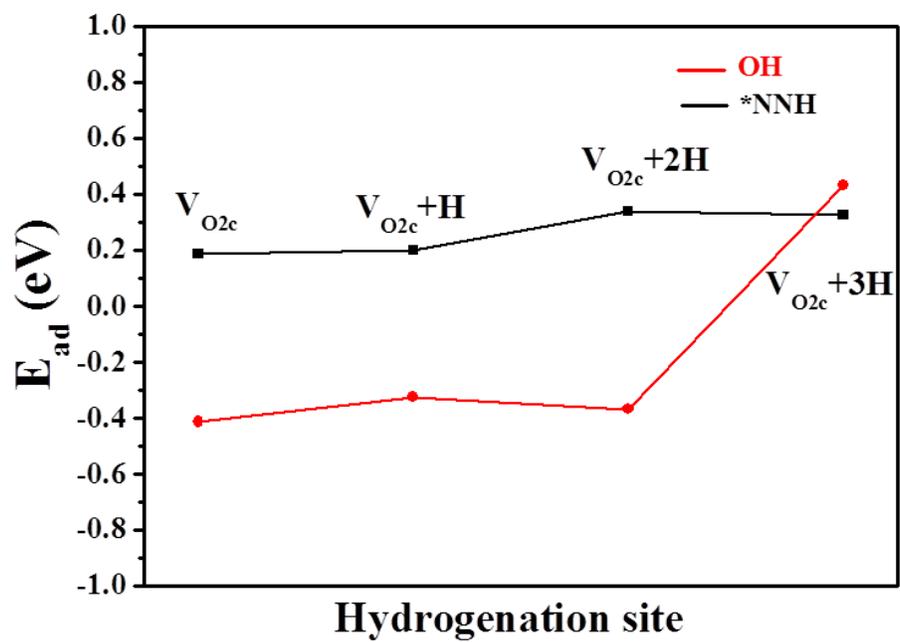

**Figure S5.** The first hydrogenation process on V$_{O2c}$-TiO$_2$ (101) surface.

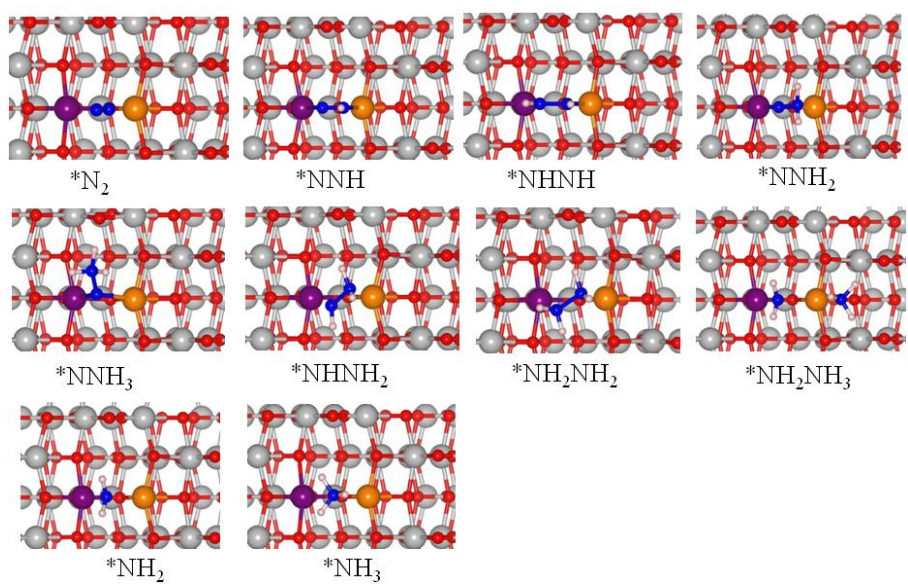

**Figure S6.** Atom configurations of NRR on $V_{O2c}$-$TiO_2$ (101) surface.

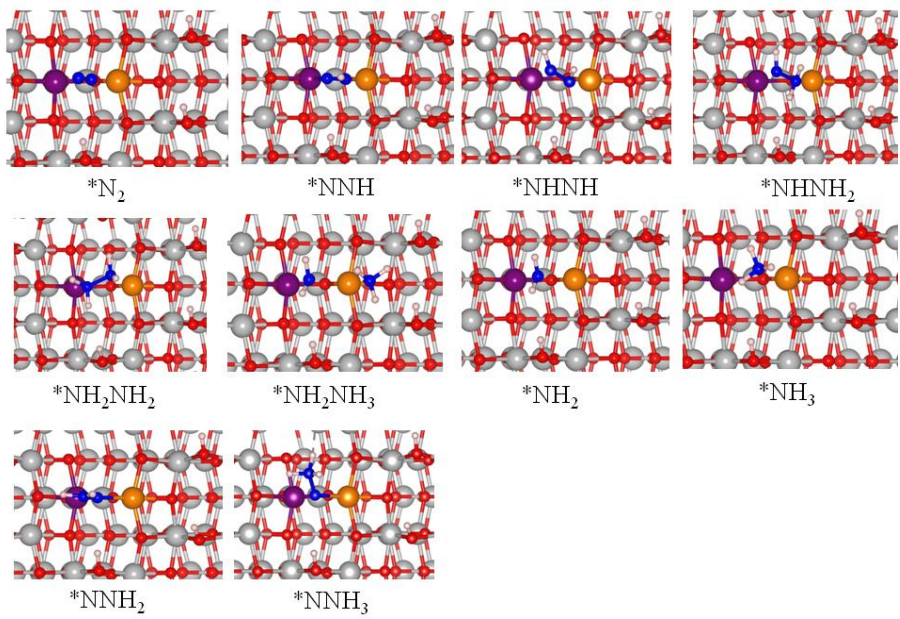

**Figure S7.** Atom configurations of NRR on $V_{O2c}$+3H-TiO$_2$ (101) surface.

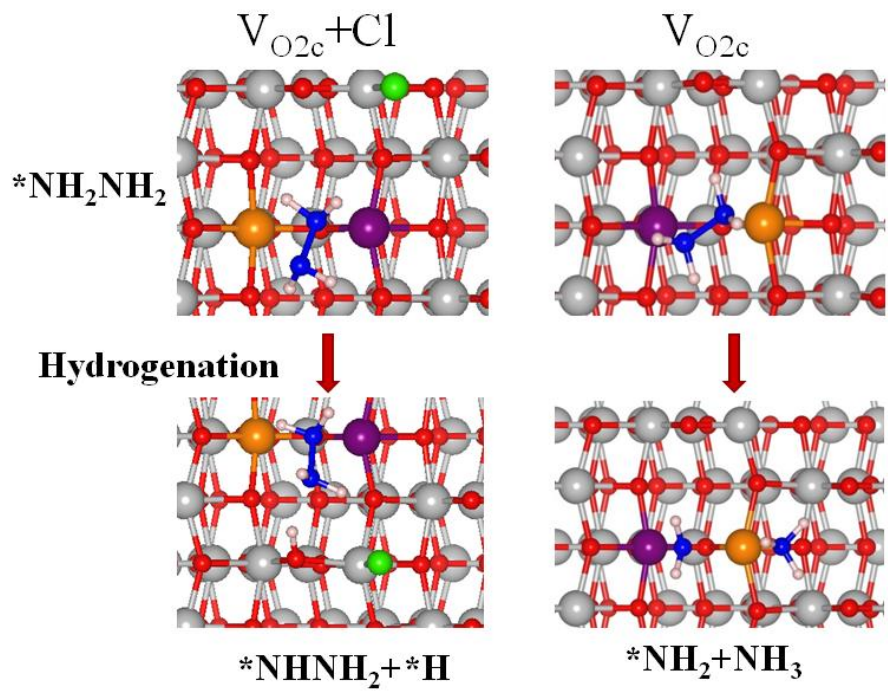

**Figure S8.** The hydrogenation process of *NH$_2$NH$_2$ on V$_{O2c}$+Cl- and V$_{O2c}$+Ti$_{int}$-TiO$_2$ (101) surfaces.

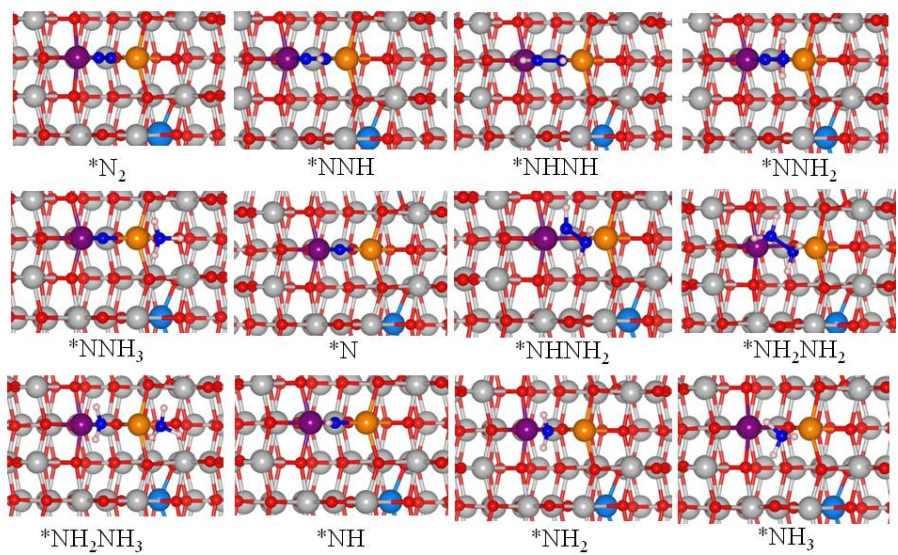

**Figure S9.** Atom configurations of NRR on $V_{O2c}+Ti_{int}$-TiO$_2$ (101) surface.